\documentclass[useAMS,usenatbib,usegraphicx]{mn2e}
\usepackage{graphicx}
\usepackage{here}

\begin{document}

\newcommand{\Sauron}{{\tt SAURON}}
\newcommand{\XSauron}{{\tt XS}auron}
\newcommand{\tiger}{{\tt TIGER}}
\newcommand{\rotcur}{{\tt ROTCUR}}
\newcommand{\gipsy}{{\tt GIPSY}}
\newcommand{\SN}{$S/N$}
\newcommand{\AN}{$A/N$}
\newcommand{\Ha}{H$\alpha$}
\newcommand{\Hb}{H$\beta$}
\newcommand{\Hg}{H$\gamma$}
\newcommand{\lda}{$\lambda$}
\newcommand{\mgb}{Mg$b$}
\newcommand{\mgt}{Mg$_2$}
\newcommand{\OI}{[{\sc O$\,$i}]}
\newcommand{\OII}{[{\sc O$\,$ii}]}
\newcommand{\OIII}{[{\sc O$\,$iii}]}
\newcommand{\NI}{[{\sc N$\,$i}]}
\newcommand{\NII}{[{\sc N$\,$ii}]}
\newcommand{\NeIII}{[{\sc Ne$\,$iii}]}
\newcommand{\SII}{[{\sc S$\,$ii}]}
\newcommand{\HI}{{\sc H$\,$i}}
\newcommand{\HII}{{\sc H$\,$ii}}
\newcommand{\plm}{$\, \pm \,$}
\newcommand{\Vsys}{$V_\mathrm{sys}$}
\newcommand{\Vrot}{$V_\mathrm{rot}$}
\newcommand{\Vrad}{$V_\mathrm{rad}$}
\newcommand{\epot}{$\varepsilon$}

\newcommand{\czero}{$c_0$}
\newcommand{\sone}{$s_1$}
\newcommand{\cone}{$c_1$}
\newcommand{\stwo}{$s_2$}
\newcommand{\ctwo}{$c_2$}
\newcommand{\sthree}{$s_3$}
\newcommand{\cthree}{$c_3$}

\def\kms{$\mbox{km s}^{-1}$}
\def\Myr{$\mbox{M}_\odot\mbox{ yr}^{-1}$}
\def\etal{et al.~}
\def\deg{^\circ}
\def\asim{\mathord{\sim}}
\def\farcs{\hbox{$.\!\!^{\prime\prime}$}}

\def\JF#1{{\bf {\sc JFB Comment:} #1}}
\newcommand{\refer}{({\sc References}) }
\newcommand{\KF}{{\sc KF Comment: }}
\newcommand{\EE}[1]{{\sc EE Comment: #1 }}

\title[A Bar Signature and Central Disk in the Gaseous and Stellar Velocity Fields of NGC~5448]
{A Bar Signature and Central Disk in the Gaseous and Stellar Velocity Fields of NGC~5448}
\author[Kambiz Fathi et al.]{Kambiz Fathi$^{1,2}$\thanks{E-mail: kambiz@cis.rit.edu};  
Glenn van de Ven$^{3}$; 
Reynier F. Peletier$^{1,4}$;
Eric Emsellem$^{5}$; \newauthor
Jes\'us Falc\'on-Barroso$^{3}$;
Michele Cappellari$^{3}$; 
Tim de Zeeuw$^{3}$.\\ \ \\ 
$^1$Kapteyn Astronomical Institute, Postbus 800, 9700 AV, Groningen, The Netherlands \\
$^2$Rochester Institute of Technology, 85 Lomb Memorial Drive, Rochester, New York 14623, USA\\
$^3$Sterrewacht Leiden, Niels Bohrweg~2, 2333~CA Leiden, The Netherlands \\
$^4$School of Physics \& Astronomy, University of Nottingham, Nottingham, NG7 2RD, UK\\ 
$^5$Centre de Recherche Astronomique de Lyon, F-69561 Saint Genis Laval Cedex, France} 

\date{submitted / accepted }
\pagerange{\pageref{firstpage}--\pageref{lastpage}} \pubyear{2005}
\maketitle
\label{firstpage}

\begin{abstract}
We analyse \Sauron\ kinematic maps of the inner kpc of the 
early-type (Sa) barred spiral galaxy NGC~5448. 
The observed morphology and kinematics of the emission-line gas
is patchy and perturbed, indicating clear departures from circular 
motion. The kinematics of the stars is more regular, and display 
a small inner disk-like system embedded in a large-scale rotating 
structure. We focus on the \OIII\ gas, and use a harmonic 
decomposition formalism to analyse the gas velocity field. 
The higher-order harmonic terms and the main kinematic features 
of the observed data are consistent with an analytically constructed 
simple bar model. The bar model is derived using linear theory, considering 
an $m=2$ perturbation mode, and with bar parameters which are consistent 
with the large-scale bar detected via imaging. We also study optical 
and near infra-red images to reveal the asymmetric extinction in NGC~5448, 
and we recognise that some of the deviations between the data and the
analytical bar model may be due to these complex dust features. Our study 
illustrates how the harmonic decomposition formalism can be used as a powerful 
tool to quantify non-circular motions in observed gas velocity fields.
\end{abstract}

\begin{keywords}
galaxies: bulges -- galaxies: spiral --
galaxies: evolution -- galaxies: formation -- 
galaxies: kinematics and dynamics -- galaxies: structure
\end{keywords}

\section{Introduction}
\label{sec:5448intro}
Dynamical studies of spiral galaxies often make the distinction 
between the bulge and the disk: the bulge is associated with the 
hot ``spheroidal'' component which mainly contains stars, and the 
disk is a rapidly rotating thin structure which also contains a 
substantial fraction of the total amount of interstellar gas. 
Line-of-sight velocity 
distributions are efficient probes of the dynamical structures
of these systems, and can be used to derive the mass distribution, 
structural properties, and perturbations of the gravitational 
potential linked to $m=1$ or $m=2$ modes, or to external triggers 
such as interactions. These can be obtained with classical long-slit 
spectrography, although it requires a-priori assumptions for the 
orientation of the slits.
A time-consuming way to obtain velocity fields through single slits 
is to spatially scan the galaxy with successive individual exposures
(e.g., Statler 1994; Ohtani 1995). 
Integral field spectrography in the optical 
(IFS, Adam \etal 1989; Afanasiev \etal 1990; Bacon \etal 1995; 
Davies \etal 1997; Garc\'{\i}a-Lorenzo, Arribas \& Mediavilla 2000;
Roth, Laux \& Heilemann 2000; Bacon \etal 2001; Bershady \etal 2004) 
was designed to uniformly cover the field of view which makes it possible 
to study these systems in a much improved way: the analysis is not 
constrained by a priori choices of slit position, and the gaseous 
and stellar components can be observed simultaneously. 
Optical IFS is often used complementarily to \HI\ and Fabry-Perot 
scanning interferometers, (e.g., van Gorkom \etal 1986; 
Plana \& Boulesteix 1996) as the latter usually have excellent 
spatial coverage but a limited spectral domain,
only probing a couple of lines (mostly in emission).

Although circular rotation is the dominant kinematic feature of the 
disk component, previous observational as well as theoretical studies 
have shown the presence of non-circular motions
(e.g., Freeman 1965; Combes \& Gerin 1985; 
Shlosman, Frank \& Begelman 1989; Athanassoula 1994; 
Shlosman \& Noguchi 1993; Moiseev 2000; 
Regan \& Teuben 2004; Wong, Blitz \& Bosma 2004). 
Cold disk-like systems are more prone than bulges to 
perturbations and instabilities, the dissipative gaseous 
component being more responsive than the stellar component 
(e.g., Thielheim \& Wolff 1982). Perturbations of the gas dynamics 
can lead to significant galaxy evolution via, e.g., redistribution 
of the angular momentum, triggering of star formation, or building 
of a central mass concentration. 

Bars are potential actors in this context as they can drive mass 
inwards or outwards, and may participate in the complex process 
which eventually leads to the fuelling of an active nucleus
(e.g., Simkin, Su \& Schwarz 1980, Sakamoto, Baker \& Scoville 2000). 
The presence of large-scale bars has been correlated with starbursts 
(e.g., Martinet \& Friedli 1997) and nuclear rings (Shlosman 1999), 
but only weakly if at all 
with nuclear activity (e.g., Knapen \etal 2000). This may not be so 
surprising as the fuelling of an active galactic nucleus (AGN), 
i.e., gas accretion onto a massive black hole, involves 
rather small spatial and short time-scales, 
so might not be related to the kpc scale bar. 
Inner small bars (Shlosman \etal 1989) may sometimes help
to link the different spatial scales (Emsellem, Goudfrooij \& Ferruit 2003), 
but they probably cannot serve as a universal mechanism. 
AGN activity is short-lived and occurs 
in the central astronomical units of galaxies, whereas current 
studies have focused on long-lived phenomena influencing the 
central kpc scale. 
In order to properly study the influence of a bar on the redistribution
of mass in the central kpc of a galaxy, we need to obtain constraints 
both on the source of the underlying gravitational potential 
and on some tracer of the on-going perturbations.
This can be achieved by simultaneously studying 
the stellar and gas dynamics.

We have observed 24 early-type spiral (Sa) galaxies with the \Sauron\ 
Integral Field Spectrograph (Bacon \etal 2001), mounted at the 4.2m 
William Herschel Telescope of the Observatorio del Roque de los 
Muchachos at La Palma (de Zeeuw \etal 2002). 
Most of these galaxies are found to have gas velocity fields that strongly 
deviate from that of a simple rotating disk. For the present study, we 
selected NGC~5448 out of the Sa \Sauron\ sample which shows a clear 
sign of the presence of a bar in its photometry and which has a 
significant amount of ionised-gas (Falc\'on-Barroso \etal 2005). 
In this paper, we analyse both the stellar and gas kinematics obtained 
with \Sauron, and quantify the non-circular gaseous motions in NGC~5448.
To this aim, we build models with elliptical streaming motion and 
compare the resulting velocity structures with our observational data, 
using a harmonic decomposition technique. In section~\ref{sec:data}, 
we present our data. Section~\ref{sec:analysis} outlines our analysis 
method of the gas kinematics. In section~\ref{sec:N5448bar}, we build a 
bar model for NGC~5448, and we present and discuss the corresponding results 
in sections \ref{sec:results} and \ref{sec:conclusions}.

\section{The Data}
\label{sec:data}
NGC~5448 is a barred Sa active galaxy with prominent irregular 
dust lanes at different spatial scales.
The RC3 classification of NGC~5448 is (R)SAB(r)a, with 
inclination of $64\deg$, systemic velocity of 1971 \kms, 
position angle (PA) of $115\deg$, and total $B$-band magnitude of 11.93.
Photometry shows that this galaxy hosts a large-scale bar with 
two well-defined spiral arms emerging near the ends of the 
bar (Eskridge \etal 2002). The inner parts of the arms are 
somewhat patchy and form a broken ring, the outer parts of 
the arms being smoother (Sandage \& Bedke 1994). This galaxy 
has a nuclear elongated feature of 
about 10\arcsec\  (Fathi \& Peletier 2003) with bluer 
$V-H$ colour than its surroundings (Carollo \etal 2002). 
Ho, Filippenko, \& Sargent (1995) classified the nucleus 
as an AGN: its central aperture spectrum exhibits a clear 
broad-line emitting region. Laine \etal (2002) 
classified NGC~5448 as a non-Seyfert.

\subsection{\Sauron\ Observations}
\label{sec:sauron}
We observed NGC~5448 with \Sauron\ on April the fourteenth 2004. 
Detailed specifications for the instrument, data, 
reduction procedure, and the data preparation procedure can be found 
in Bacon \etal (2001), de Zeeuw \etal (2002), and Emsellem \etal (2004). 
A brief summary of the instrument and data characteristics 
is presented in Table~\ref{tab:data}. 
We obtained 4 exposures of 1800 s each, 
producing a total of 1431 simultaneous galaxy spectra per frame, 
together with 146 sky spectra 1\farcm9 away from the main field.
We also observed standard stars to be used for accurate 
velocity, flux, and line-strength calibration of our observed galaxy. 
Wavelength calibration was done using arc lamp exposures taken before 
and after each target exposure.
The standard reduction was performed using the \XSauron\ software 
package developed at the CRAL (Lyon), providing a fully calibrated 
datacube for each individual exposure. We merged the 4 spatially 
dithered exposures after a resampling to a common spatial scale 
of $0\farcs8 \times 0\farcs8$ per pixel, leading to a total of 1973 
spectra within our field.
\begin{center}
\begin{table}
\caption{\Sauron\ data characteristics.}
\tabcolsep=7.pt
\small \begin{tabular}{ll}
\hline 
Field of View 		& $41\arcsec \times 33\arcsec$	\\
Pixel Size		& $0\farcs8$			\\
Instrumental Disp.	& 108 \kms			\\
Spectral Range		& [4820 - 5280] \AA		\\
Spectral Features	& \Hb, \OIII, Fe5015, \mgb, \NI, Fe5270\\
\hline
\end{tabular}
\label{tab:data}
\end{table}
\end{center}

\begin{figure*}
\center \includegraphics[scale=2.30,trim=0.cm 0.cm 0cm 0cm]{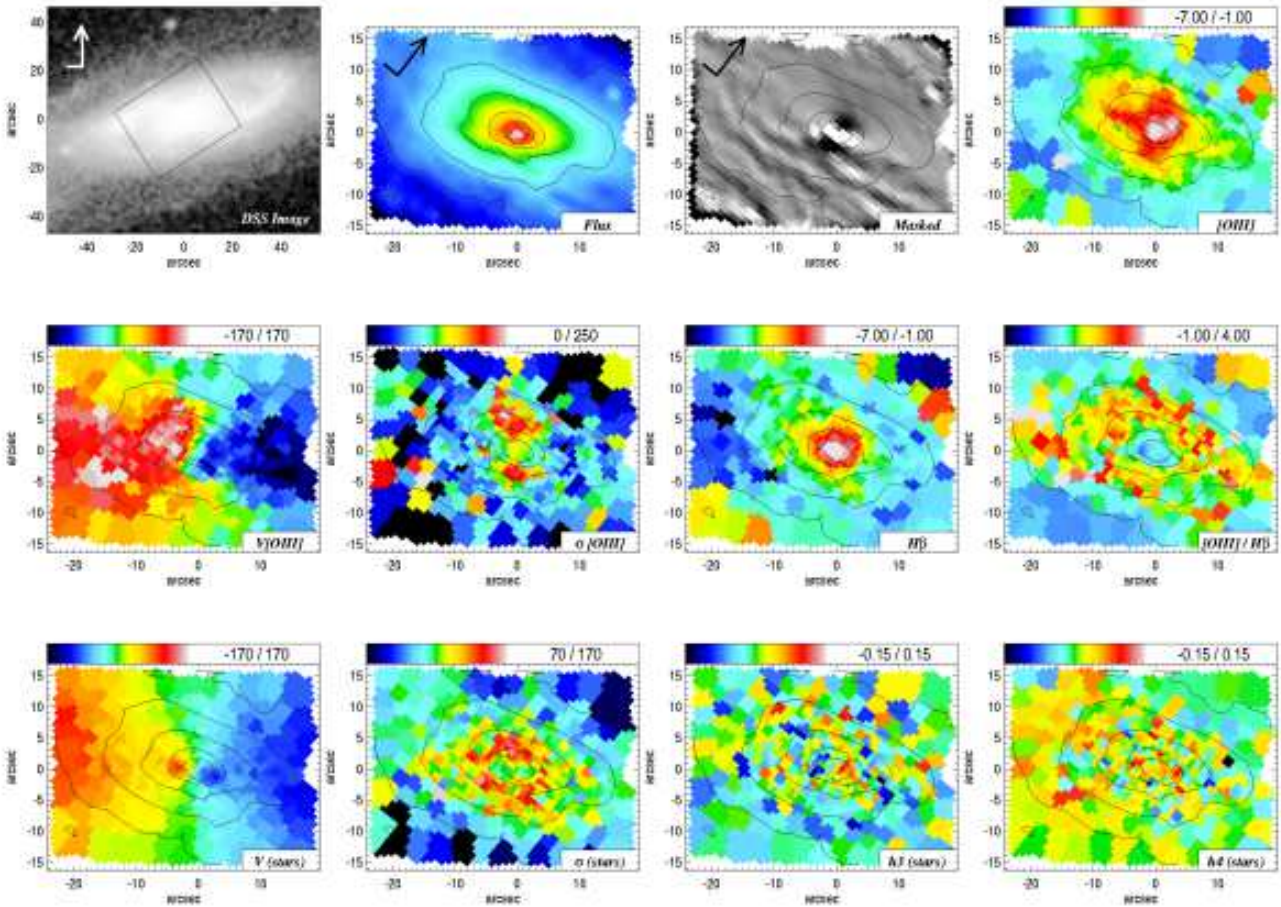} 
\caption{Top Left: Digitised Sky Survey image with \Sauron\ footprint and
north-east orientation arrow. All other panels show the \Sauron\ data.
The stellar flux map and unsharp-masked \Sauron\ image are given in mag 
arcsec$^{-2}$ with arbitrary zero point, and north-east direction as 
indicated. The titles are indicated at the bottom right corner of each panel 
with plotting ranges according to the top color bar. All \Sauron\ maps have
the same orientation with overplotted stellar contours in magnitude steps of
0.25, and all velocities and velocity dispersions are given in \kms. }
\label{fig:datamaps}
\end{figure*}

\subsection{Stellar and Gaseous Kinematic Maps}
\label{sec:starsandgas}
The signal-to-noise (\SN) ratio of individual spectra varies considerably 
throughout the field: in order to homogenise the \SN\, we applied the 
Voronoi 2D-binning technique of Cappellari \& Copin (2003). We have 
set a target \SN\ value of 60 to be able to derive reliable higher-order 
moments of the stellar velocity distribution. Penalised Pixel Fitting 
(pPXF, Cappellari \& Emsellem 2004) was used to derive 
the stellar line-of-sight velocity distribution parametrised by 
a Gauss-Hermite series up to and  including the fourth order 
(van der Marel \& Franx 1993; Gerhard 1993). The optimal stellar 
template in pPXF was built with a combination of stellar
spectra from Jones (1997) and synthesis spectra from 
Vazdekis (1999) as in Emsellem \etal (2004). The derived stellar 
mean velocity $V$, mean velocity dispersion $\sigma$, and higher 
order velocity moments $h_3$ and $h_4$ maps, are presented in 
bottom row of Fig.~\ref{fig:datamaps}. We discuss the maps in 
section \ref{sec:results}.

The method which we have applied to derive the emission-line
kinematics is described and validated in Sarzi et al.\ (2005) and
Falc\'on-Barroso et al.\ (2005). It consists of iteratively
searching for the emission-line velocities and velocity
dispersions, while solving at each step for both their amplitudes
and the optimal combination of the stellar templates over the
full \Sauron\ wavelength range. No masking of the regions affected
by emission is thus required.

Our binning scheme was optimised for the 
derivation of reliable stellar kinematics, but is not necessarily 
well adapted to analyse the emission-line gas. Since the
gaseous component commonly rotates faster than the stellar
component, to obtain reliable gas kinematics, in general a lower
\SN\ is required. Moreover, gas and stars do not necessarily share
the same spatial distributions. As a consequence, the stellar bins
are often larger than necessary for the gas, implying a loss of
spatial information for analysing the gaseous component (Fathi 2004). 
Accordingly, the data that we present in this paper has a minimum 
stellar \SN\ of 60 and a minimum emission-line amplitude-over-noise of 3.

Finally, we have carefully inspected the emission-line profiles
to search for possible asymmetric profiles but have not found any
significant deviation from a pure Gaussian, nor any significant
indication for more than one component being present in the
profiles, at least at the instrumental resolution of \Sauron. 
We know from other \Sauron\ studies of e.g., NGC~1068
(Emsellem et al.\ 2005) that we can resolve several components 
in very strong AGNs.

\section{Analysing Gas Velocity Fields}
\label{sec:analysis}

\subsection{Circular and Non-Circular Kinematics}
\label{sec:kinematics}
In this section we start with a simple mathematical model, and then 
add complexity as required by the observational constraints.
In the simplest model, the galactic disk is purely rotational, 
has negligible velocity dispersion, and is infinitesimally thin. 
In the presence of axisymmetric radial and vertical velocities, and 
when taking into account the effect of projection and the convention that 
positive line-of-sight velocities correspond to recession, the 
line-of-sight velocities $V_\mathrm{los}$ can be represented by
\begin{equation}
\label{eq:Vxy}
\begin{array}{rl}
V_\mathrm{los}(R,\psi, i) = V_\mathrm{sys} &\hskip -3mm + V_\mathrm{rot}(R) \cos\psi \sin i \\[2mm]
			     &\hskip -3mm + V_\mathrm{rad}(R) \sin\psi \sin i \\[2mm]
			     &\hskip -3mm + V_z(R) \cos i,
\end{array}
\end{equation}
where \Vsys\ is the systemic velocity of the galaxy, \Vrot\ and \Vrad\ are
the rotational and radial velocities, and $V_z$
is the vertical velocity component, which we set to zero
throughout this paper. The inclination $i$ of the galaxy ranges
from $i=0\deg$ for a face-on viewing and $i=90\deg$ for an edge-on
viewing. Furthermore, $(R,\psi)$ are polar coordinates in the 
plane of the galaxy related to observable Cartesian coordinates 
$x$ and $y$ (in the plane of the sky) by
\begin{equation}
\label{eq:sincos}
\left\{ \begin{array}{l}
\cos\psi = \displaystyle\frac{-(x-X_{\rm cen})\sin\phi_0 + (y-Y_{\rm cen})\cos\phi_0}{R},\\[2mm]
\sin\psi = \displaystyle\frac{-(x-X_{\rm cen})\cos\phi_0 - (y-Y_{\rm cen})\sin\phi_0}{R \cos i},\\
\end{array} \right.
\end{equation}
where $X_\mathrm{\rm cen}$ and $Y_\mathrm{\rm cen}$ are the 
coordinates for the centre, and $\phi_0$ is the PA of the
projected major axis of the disk, measured with respect to north in 
counterclockwise direction.
This simple model cannot explain most observed velocity fields. 
The gas kinematics in real galaxies exhibits radial and/or vertical 
motions due to, e.g., the presence of bars, spiral arms, which create
angle-dependent velocities which cannot be explained using Eq.~(\ref{eq:Vxy}), 
and hence additional ingredients in the analysis method are required.

Several attempts have been made to investigate more complex
velocity fields (Sakhibov \& Smirnov 1989; Canzian 1993; 
Schoenmakers, Franx \& de Zeeuw 1997 [hereafter SFdZ]; 
Fridman \& Khoruzhii 2003; Wong \etal 2004; Krajnovi\'c \etal 2005). The main idea of these 
techniques is to deal separately with the ``unperturbed'' underlying component 
and the residual velocity field, obtained by subtracting the reconstructed 
unperturbed velocity field from the data. To solve the problem, as stated, one
would have to know the unperturbed velocity component at a given radius. 
Unfortunately we do not know beforehand the unperturbed velocity field. A powerful technique for 
unveiling the perturbations makes use of the expansion of the
velocity information in Fourier harmonic components. Following
the terminology of SFdZ (see their appendix for the detailed derivation),
given that the line-of-sight velocity can be expressed as a two-dimensional 
analytic function of galactocentric coordinates, it can be written as a 
Fourier series: 
\begin{equation}
\label{eq:expansion}
V_\mathrm{los}= V_\mathrm{sys} + 
		\sum^{k}_{n=1}{\big(c_n(R)\cos\,n\psi + s_n(R)\sin\,n\psi\big)\sin i}, 
\end{equation}
where $k$ is the number of harmonics, 
and $c_n$ and $s_n$ give us information about the nature of the 
perturbations, and they are tightly connected to the underlying 
potential. Note that $c_1=V_\mathrm{rot}$ and $s_1=V_\mathrm{rad}$.
Furthermore, numerical simulations by Burlak \etal (2000) have 
shown that this formalism is very stable and its results are 
rather insensitive to the presence of holes in the velocity field.

\subsection{Tilted-Ring and Harmonic Decomposition}
\label{sec:decomposition}
\begin{figure*}
\center \includegraphics[scale=2.30,trim=0.cm 0.cm 0.cm 0.cm]{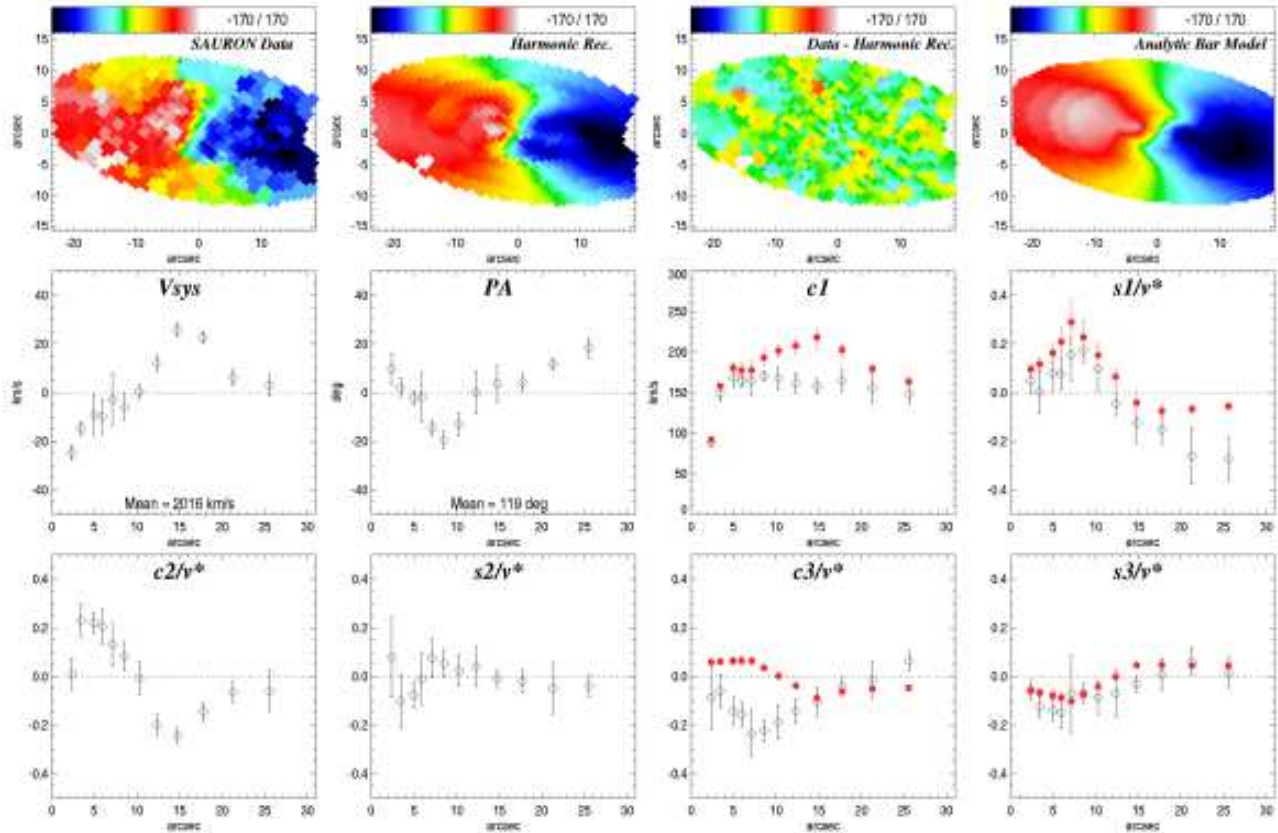} 
\caption{Top row, from left to right: observed \Sauron\
gas velocity field of NGC~5448; reconstruction based on the
harmonic decomposition of the \Sauron\ gas velocity field;
residual field (data - harmonic reconstruction); and the analytic
bar model, which reproduces best the main kinematic features of
the observed gas velocity field. Middle and bottom row: the
harmonic parameters as a function of outer radius of each ring,
and $v^* =c_1 \sin i$. The over-plotted red filled circles are the
analytically calculated first and third harmonic terms for the bar
model (the second terms are zero by construction), with the error
bars corresponding to the $3\sigma$ (99.7\%) confidence level (see
section 4 for details). The orientation of the maps is the same as
in Fig. \ref{fig:datamaps}, and the mean PA value is given in the
north-east direction.}
\label{fig:analysis}
\end{figure*}

Our quantitative analysis uses the formalism of 
Eq.~(\ref{eq:expansion}) combined with the tilted-ring method 
(Rogstad, Lockart \& Wright 1974; Rogstad, Wright \& Lockart 1976), 
inspired from the \rotcur\ routine in the \gipsy\ package 
(Begeman 1987; van der Hulst \etal 1992). 
Accordingly, we divide the galaxy disk into concentric ellipses 
within which we fit the rotation velocity and the set of
geometric parameters of Eq.~(\ref{eq:expansion}). 
The method assumes that each pixel in the velocity field is 
identified with a unique position in the galaxy and that circular 
rotation is the dominant feature. We assume that our measurements 
refer to positions on a single inclined disk, i.e., that we do not
expect a strong warp and/or overlapping spiral arms within the observed field. 
Accounting for harmonic perturbations up to order 3, 
this method results in a large parameter-space problem,
and it requires some additional assumptions to obtain a physically 
meaningful solution. We obtain the parameters by proceeding as follows: 

\begin{enumerate}
\item We start from a simple inclined two-dimensional disk. 
In case of erroneous inclination, one would detect systematics 
in the residual field as explained in section~\ref{sec:errors}.
Using a two-dimensional disk produces a robust model, in the sense 
that warps and other deviations from this basic model should 
be prominent in the residual maps. 
\item Secondly, we fix the dynamical centre of the galaxy to coincide 
with the photometric centre. The photometric center is obtained by fixing the
center of the $H$ band image, since the $H$-band image is less affected by dust. 
The HST $V$-band image and the \Sauron\ image are then aligned with the HST $H$-band image.
\item At every radius, we fit $V_{\rm los} = c_0 + c_1 \cos \psi\sin i$, 
varying $c_0$, PA, and $c_1$. The result is used 
to fix the systemic velocity (\Vsys$=c_0$) simply as the mean value over radius, 
using Tukey's bi-weight mean formalism (Mosteller \& Tukey 1977), 
which is particularly advantageous for being non-sensitive to 
outliers. After \Vsys, we fit the PA and $c_1$, followed by 
fixing the PA in the same way as the \Vsys\ was fixed.
\item For each ring, we import the values derived according 
to the recipe above, and fit the rotation curve and the higher 
harmonic components (up to order 3) applying a $\chi^2$-minimisation 
scheme. These results are presented in Fig.~\ref{fig:analysis}.
\end{enumerate}

Analysing the first, second, and third harmonic components only, is an
efficient way to analyse specific elements of the perturbations on the 
underlying potential. A perturbation of order $m$ creates $m-1$ and $m+1$ 
line-of-sight velocity terms (see e.g., Canzian 1993 and SFdZ). 
As a result, our fitted \cone, \sone, \cthree, and \sthree\ terms 
contain information about possible $m=2$ perturbations. 

When sectioning the field into concentric
rings, the radial extent of each ring is a 
free parameter which has to be pre-determined. 
For a range of possible ring radii, 
we have examined the fitted velocity information. 
Determination of the radial thickness of the rings becomes a 
trade-off between smoothness of the fitted velocity field, 
and robustness of the fit. Radially thin rings result in smooth 
fits, and thus smooth residual (Data - Fit) fields, whereas 
radially thick rings result in lower errors for the fitted parameters. 
Patchy gas distribution causes non-uniform distribution of points, 
and it is important to make sure 
that all the rings include enough points to fit harmonic parameters 
to reasonable accuracy. This is particularly important in the innermost 
and outermost points. We adopt a geometric increase of the ring radii
(meaning that the radial width of the rings is increased by a 
factor 1 + step). Here, we adopt a step of $0.2$, and only fit the inner
30\arcsec\ of the \Sauron\ velocity field, since rings larger than this radius 
are not sufficiently covered and hence do not deliver reliably derived
parameters. Finally, since we are simultaneously fitting 6 harmonic components, 
we make sure that each ellipse contains at least $25$ data points, in order to 
obtain reliable errors.

\subsection{Errors}
\label{sec:errors}
Warner \etal (1973) and van der Kruit \& Allen (1978) showed that assuming 
wrong input disk parameters could cause recognisable signatures in the 
residual velocity fields. SFdZ quantified these signatures in terms of harmonic
expansion, and showed that the Fourier components due to non-circular motions 
will mix with those due to erroneous disk parameters 
$X_\mathrm{cen}, Y_\mathrm{cen}, i,$ and PA. 
Errors in the kinematic centre translate to the \czero, \ctwo\ and 
\stwo\ terms, while PA errors appear in the \sone\ and \sthree\ 
terms, and inclination errors in the \cone\ and \cthree\ terms. 
In the case of erroneous kinematic centre, the \czero\ and the \ctwo\ 
terms fall off as $\frac{1}{R}$ whereas the presence of a radius-independent 
$m=1$ mode shows the relation $c_0 \sim 3c_2$ (Schoenmakers 1999). 
Our chosen centre coordinates do not deliver any of these situations. 
Another source of error is ``pixel sampling'' for which errors propagate 
onto the \cone\ (as the rotation curve rises more gradually) and the \cthree\ term
(as if inclination changes). This effect may also propagate onto the \sone\ 
and \sthree\ parameters, although not significantly (Wong 2000). 
Our data and analysis does not require consideration of this effect since 
our spatial resolution is high enough.

Our data extraction code delivers error measurements for each 
individual pixel (Falc\'on-Barroso et al. 2005). 
We calculate the errors for the tilted-ring and harmonic parameters 
by means of Monte Carlo simulations. Repeated 
application of the tilted-ring method to the Gaussian randomised gas 
velocity field yields the uncertainties on the harmonic parameters.
Our 500 simulations show that choosing geometrically increasing ring radii
with steps of $0.2$ indeed yields satisfactory errors. The errors are low, 
and the parameter profiles are represented by an adequate number 
of points.

\section{A Bar Model for NGC~5448}
\label{sec:N5448bar}
The SFdZ formalism is a very powerful tool for analysing velocity
fields. One can analytically derive the higher-order harmonic
parameters for potentials of desired form. Schoenmakers (1999)
derived the harmonic parameters for a lopsided potential, and a
more extensive expansion was done by Wong \etal (2004), who
derived the higher-order harmonics for simple bar and two-armed
spiral perturbations.

We import their model for a globally elongated gravitational
potential, which, in the plane of the galaxy, consists of
an axisymmetric component $\Phi_0(R)$ from Kuijken \& Tremaine
(1994), with a bi-symmetric potential of the form $\Phi_2(R) \cos
2\theta$ added to it. In the case of flat rotation curve, the
potential is written as
\begin{equation}
\label{eq:barpot}
\left\{ \begin{array}{l}
\Phi_0 (R) = v_{c}^2 \ln R,\\[2mm]
\Phi_2 (R) = -\varepsilon v_{c}^2 \; / \; 2,\\
\end{array} \right.
\end{equation}
where $v_{c}$  is the circular velocity and \epot\ is the
elongation of the potential. Introducing a constant damping term
into the equations of motion and considering only the $m=2$
perturbation term (Lindblad \& Lindblad 1994; Wada 1994), linear
epicyclic approximation of Lin \& Shu (1964) was applied to obtain
an analytical solution for gaseous orbits in this barred
potential. Since this model is restricted to $m=2$ perturbation,
it affects only the first and third harmonic parameters (\cone,
\sone, \cthree, and \sthree). These depend on the amplitude of the
damping term ($\lambda$), the ellipticity of the potential
(\epot), the corotation radius (CR), and the viewing angle
($\theta$), where $\theta=0$ corresponds to end-on view.

We use these analytic harmonic parameters to reconstruct
the signatures of prominent non-circular motions in the observed
velocity fields. We build a library of models with varying bar
parameters ($0<\varepsilon<0.5$, $0<\lambda<0.5$ and a range of
bar sizes), viewed from different angles
($0\deg\leq\theta\leq180\deg$). The effects of $\lambda$ and
\epot\ on the velocity field are predictable. The bar signature is
weaker for a larger damping factor and/or a smaller flattening of
the potential. However, the effect of varying $\theta$ is not
straightforward.

Investigating our library of dissipative bar models, we
find that only for a certain combination of parameters the model
exhibits a strong twist of the gas zero-velocity curve similar to
our data (Fig.~\ref{fig:datamaps}). We quantify the comparison by
calculating, for each bar model in our library, the discrepancy
between the reconstructed velocity field and the observed velocity
field, in terms of the goodness-of-fit parameter $\chi^2$. Care
has to be taken at the inner Lindblad resonance (ILR), since there
the linear epicyclic theory breaks down. At the ILR, due to the presence 
of non-linear terms, which are not accounted for by the analytic epicyclic 
theory, the weak-bar model is not expected to reproduce the velocity field.
Hence, for a small region around the ILR, we interpolate the bar models. 

The minimum goodness-of-fit parameter
$\chi^2_\mathrm{min}$ yields the best-fit bar model, for which we
show in Fig.~\ref{fig:analysis} the reconstructed velocity field
and harmonic terms (red filled circles). The best-fit model
accounts for an $m=2$ perturbation of a single bar, with potential
ellipticity $\varepsilon = 0.15 \pm 0.02$, damping term $\lambda =
0.12 \pm 0.03$, viewing angle $\theta = 19\pm 3 \deg$, and CR =
$37 \pm 4 \arcsec$. The $3\sigma$ errors on the bar parameters
follow from the bar models for which the difference between the
corresponding $\chi^2$ and $\chi^2_\mathrm{min}$ is below the 99.7\%
level. From the harmonic terms of these bar models, we also obtain
an estimate on the uncertainty in $c_1$, $s_1$, $c_3$ and $s_3$,
indicated by the red error bars in Fig.~\ref{fig:analysis}. Given
the significant second harmonic term and the simplicity of the
analytic bar model, it is not surprising that it cannot provide a
perfect fit to the data. Still, this generic bar model does
reproduce the main features in the observed velocity field, as
well as the overall behaviour of the first and third harmonic
terms. This supports the case of a bar as the main driver behind
the observed non-circular motions.

\section{Results} 
\label{sec:results}
The gas velocity field decomposition of section \ref{sec:decomposition} 
provides the radial profiles for the primary disk parameters and the higher
order harmonic terms. Assuming the inclination $i=64\deg$ from the RC3 catalogue
(de Vaucouleurs \etal 1991), we apply this method to our observed gas 
velocity field in an iterative manner by first varying the \Vsys\ and PA.
We then fix these parameter to their mean values and we proceed by deriving 
the \Vrot\ and higher harmonic terms (see Fig.~\ref{fig:analysis}). We detect a 
varying zeroth harmonic term (\Vsys) with variation amplitude up to 
40 \kms\ in the central 10\arcsec. We find a strong PA twist of about 
$30\deg$ in the central 10\arcsec, and outside this
radius the PA goes back to the central values. The circular velocity
component \cone\ rises steeply and peaks at 5\arcsec\ radius, followed by a 
very slow decline out to the outer radii. 
We find that the behaviour of the \sone\ term could be compared with the 
\Vsys\ or the PA: the \sone\ term changes sign at around 12\arcsec,
and at the same galactocentric radius, the \Vsys\ and the PA
change sign around their mean values. The \sone\ exhibits amplitudes of order of 
$0.2\, c_1 \sin i$. 
The bottom row in Fig.~\ref{fig:analysis} shows that, apart from the \stwo\ term, 
the higher harmonic terms significantly deviate
from zero, and indeed account for the considerable deviation of the
velocity field of NGC~5448 from pure rotation.

In the following sections, we describe the distinct features that we find in 
our kinematic maps.

\subsection{The Stellar Component} 
\label{sec:stars} 
\begin{figure*}
\center \includegraphics[scale=2.3,trim=0.cm 0.cm 0.cm 0.cm, angle=0]{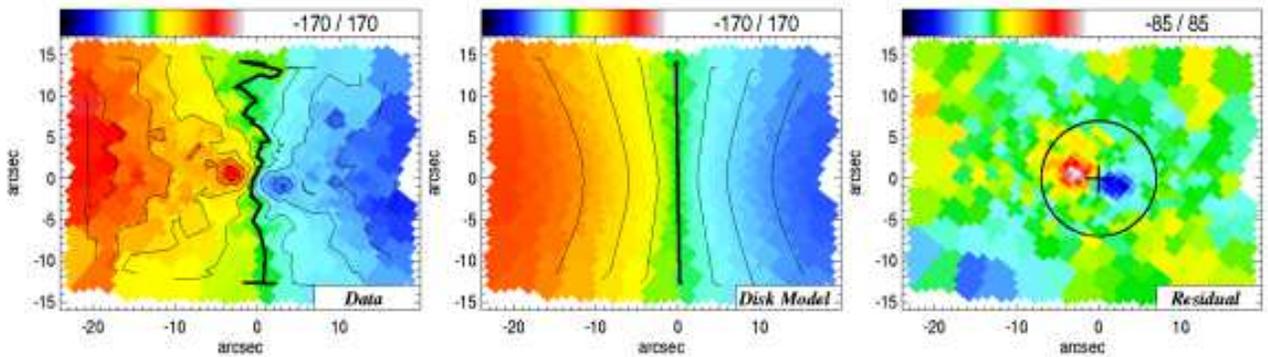} 
\caption{A thin isothermal disk model for the stellar velocity
field of NGC~5448. The circle marks the 7\arcsec\ region within
which we find a disk-like structure. The disk model, fitted to the
field outside this region, implies for the outer disk a scale
length of 18\arcsec, \Vsys$=2002$ \kms, and PA$= 91\deg$. 
The orientation of the maps is the same as in Fig. \ref{fig:datamaps}.}
\label{fig:diskmodel}
\end{figure*}

The \Sauron\ flux map in Fig.~\ref{fig:datamaps} displays a smooth stellar 
distribution and the presence of prominent dust lanes to the south of the nucleus. 
The stellar kinematics shows a prominent regular disk-like rotation with 
a smaller inner stellar disk within the central 7\arcsec\ radius. 
The stellar velocity dispersion 
decreases towards the centre, and at the location of the maximum line-of-sight
velocities for the central structure, $h_3$ anti-correlates with $V$ which
supports the argument that the central component of NGC~5448 is a central disk.
We approximate the stellar velocity field with that of an exponential thin disk 
(Freeman 1970) to emphasize the kinematic signatures of 
the central disk. Fig.~\ref{fig:diskmodel}
shows this simple inclined disk model, where the inner stellar
disk rotates faster than the outer disk. The best-fit model for
the outer stellar disk, omitting the central 7\arcsec\ and
assuming an inclination fixed to the RC3 catalogue value, yields
\Vsys$=2002$ \kms, a disk scale length $R_d=18$, and a disk 
PA$=91\deg$. Repeating the same exercise for the region inside
7\arcsec, we find that the central disk is
misaligned with respect to the outer disk by $\sim 13\deg$.

\subsection{Gas Distribution and Kinematics} 
\label{sec:gas}
Fig.~\ref{fig:datamaps} demonstrates that along the 
strong dust lanes, the gas shows a patchy distribution, 
with an asymmetric elongation of \OIII\ gas towards 
the east as well as the galactic poles. The \Hb\ distribution 
is found to be more regular, and we have found that all kinematic 
features of the \OIII\ gas are accompanied by 
\Hb\ emitting gas. The \OIII/\Hb\ map in Fig.~\ref{fig:datamaps} displays a
prominent ring-like structure at 5\arcsec-15\arcsec, indicating high ionisation
of the gas in this region.

\begin{figure}
\includegraphics[scale=2.0,trim=0.cm 0.cm 0.cm 0.cm]{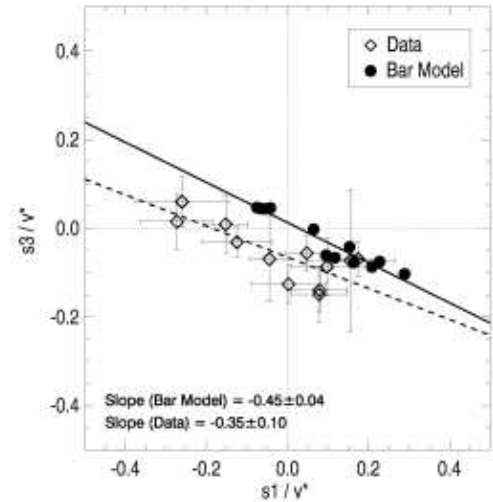} 
\caption{First and third sinusoidal harmonic terms
normalised to $v^* = c_1 \sin i$. Both the data and the bar model
show a negative slope (dashed and solid line respectively),
indicating that elliptical streaming, due to the bar potential, is
the main driver of the radial motions observed in the gas velocity
field of NGC~5448. }
\label{fig:s3s1}
\end{figure}

The gas velocity map clearly shows very prominent `S'-shaped zero-velocity 
curve with very sharp edges indicating very strong non-circular gas motions 
(Peterson \& Huntley 1980). This `S'-shaped gaseous zero-velocity curve is 
confirmed by the PA variation derived by the tilted-ring decomposition in 
Fig.~\ref{fig:analysis}. Zero-velocity curve twists can be indicative of 
strong radial motions, created by, e.g., interactions, mergers or elliptical 
streaming due to a barred potential.
Wong (2000) distinguished differences between radial motion mechanisms by 
comparing the \sthree\ versus \sone\ harmonic terms, and found that 
the behaviour of \sthree\ versus \sone\ is different between a bar model 
and a pure axisymmetric or warped model. As demonstrated in 
Wong et al. (2004; Fig.~5) the \sthree\ versus \sone\ of a warped disk 
lie on a positive slope. Moreover in case of an externally triggered 
radial flow, the points in the \sthree\ versus \sone\ graph should lie 
on a zero-slope. Setting up the same diagnostics, 
in Fig.~\ref{fig:s3s1}, we find that for our data, 
the \sthree\ versus \sone\ curve has a negative slope very similar 
to that of the bar model presented in Fig.~\ref{fig:analysis}.
This confirms the signature of elliptical streaming  
in the gas velocity field of NGC~5448.

Comparing the stellar maps with the gas maps, we find 
that the gas velocities are considerably higher: This is confirmed with 
the rotation curve presented in Fig.~\ref{fig:sauzoom}. 
We find that the gaseous $\sigma$ map displays 
some features comparable to the stellar $\sigma$ map. 
Although the stellar velocity dispersion is more steeply rising 
toward the centre, both maps show a prominent dip in the centre. 
This supports the hypothesis of the presence of a central dynamically 
cold disk-like structure, present both in the stellar and gaseous 
component. Both maps also show a dispersion increase in 
bi-polar directions just a few arcseconds from the centre. This is 
much more prominent in the gas map, as its $\sigma$ reaches values 
$\simeq 250$ \kms. This may be indicative of significant
outflows in this AGN host.

\subsection{Position Angle Values} 
\label{fig:PAs}
\begin{figure}
\center\includegraphics[scale=2.0,trim=0.1cm 0.cm 0.cm 0.cm, angle=0]{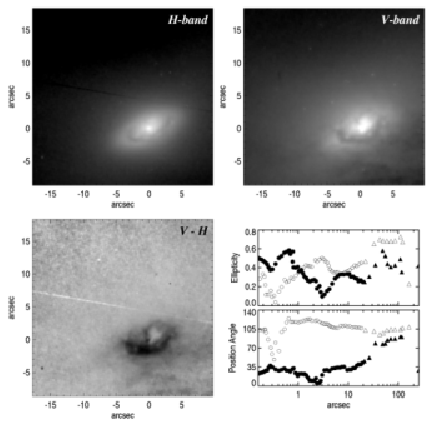} 
\caption{The top panels show HST/NICMOS $H$-band (F160W) and WFPC2
$V$-band (F606W) images of NGC~5448. The bottom left panel is the
$V-H$ map obtained by convolving the $H$-band image with the
$V$-band image PSF and vice versa, and subtracting one from the
other. All maps have pixel-size of 0\farcs0455, and the same
orientation as the DSS image with the centre marked by a cross.
The bottom right profiles are the observed (open symbols) and
deprojected (filled symbols) ellipticity and PA profiles from
Laine \etal (2002). Circles represent values derived from the
NICMOS image, and triangles show the values derived using the DSS
image. Here we only focus on the region beyond 1\arcsec.}
\label{fig:nicmos}
\end{figure}

The projected and deprojected PA profiles 
from Laine \etal (2002) in Fig.~\ref{fig:nicmos}
display strong variations in the central few arcseconds. 
The HST images display asymmetrically distributed dust within 
the central few arcsecoonds, which may cause these strong
central PA variations (also found by Kornreich \etal 2001). 
At larger radii, we find that the PA profile shows a 
variation of the order of $10\deg$, but overall it stays fairly 
constant around the nominal RC3 value, i.e. $115\deg$. 
Our tilted-ring decomposition of the gaseous velocity field, yields 
a mean gas kinematic PA of $119\deg\pm 5$, which is consistent with 
the photometric PA from RC3. In the case of the large stellar
disk, as described in section~\ref{sec:stars}, we find an offset
of $\sim 25\deg$ between the stellar kinematic PA and the
photometric mean PAs. This misalignment could be due to the effect
of the bar on the observed stellar velocity field.

\subsection{The Inner Few Arcseconds} 
\begin{figure*}
\center \includegraphics[scale=2.3,trim=0.cm 0.cm 0.cm 0.cm, angle=0]{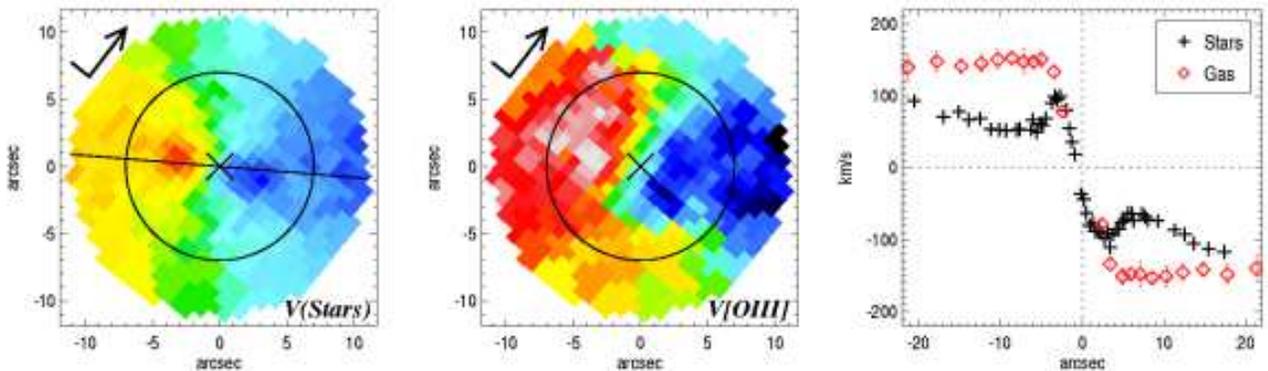} 
\caption{Zooming into the central few arcseconds of the
stellar and gaseous velocity maps of NGC~5448, using the same
velocity range as in Fig.~\ref{fig:datamaps}. Indicated are the
north-east direction (arrow), the photometric PA (straight line)
and the photometric centre (cross). The over-plotted circle
indicates the 7\arcsec\ radius for comparison with
Fig.~\ref{fig:diskmodel}. In the right panel, we present the
stellar rotation curve (extracted along the photometric PA)
together with the gas rotation curve derived from tilted-ring
decomposition.}
\label{fig:sauzoom}
\end{figure*}
In Fig.~\ref{fig:sauzoom} we zoom into the central few arcseconds of the stellar
and the gaseous velocity fields of NGC~5448 to emphasize the observed kinematic
differences between the two components. The gaseous component exhibits a
strongly twisting zero-velocity curve, and the gas rotates faster than the
stars. We find that the inner stellar disk not only is misaligned with the
outer stellar disk, but also is shifted to the south-east. 
In contrast, we find from the photometric profiles in
Fig.~\ref{fig:nicmos}, that the ellipticity of the central disk is
similar to that of the outer disk ($\sim 0.55$), as well as their
PAs. The prominent dust lanes at the south of this galaxy may explain the 
kinematic misalignment between the outer and inner stellar disks.

In the central $\sim 4\arcsec$ the gas isovelocity 
contours are comparatively straight and parallel. 
In the same region, the stellar isovelocity 
contours are indicative of the presence of a central disk. 
The tilted-ring decomposition of the gas velocity field provides the gas
rotation curve illustrated in Fig.~\ref{fig:sauzoom}. 
The gas rotates about 70 \kms\ faster than the stars in the region 
outside $\sim 4\arcsec$,  with some indication that outside this radius 
the negative velocities reach lower values than the positive 
line-of-sight velocities. This asymmetry could be caused by complex 
dust distribution or attributed to a signature of lopsidedness.

\subsection{Dust Effects or Lopsidedness?} 
\label{sec:dust}
Although the unsharp-masked \Sauron\ image shows strong signature of
dust-lanes in the region south of the nucleus, 
the central dust distribution is easier to study when looking at 
the HST images and the $V-H$ colour map in Fig.~\ref{fig:nicmos}. 
The colour map shows a strong dust lane about 1\arcsec\ at the south-east of 
the galaxy nucleus as well as an asymmetrically distributed overall dust
distribution. The larger scale WFPC2 image shows that this central 
dust lane is accompanied by several other dust lanes further away from the 
nucleus. Using the Galactic extinction law of 
Rieke \& Lebofsky (1985), the HST images yield an average extinction 
value of  $A_V \simeq 0.5$ over the central 3\arcsec, with a maximum value of
1.5.  The extinction increases significantly towards the centre, 
with a prominent peak a few arcseconds south of the nucleus (see Fig.~\ref{fig:nicmos}).

\begin{figure}
\includegraphics[scale=2.0,trim=0cm 0.cm 0.cm 0.cm]{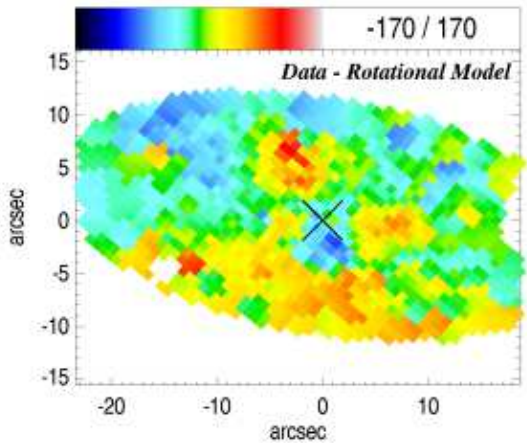} 
\caption{Residual field after subtracting the best-fit
rotational model from the observed gas velocity field. This map
reveals the strong non-circular motions in NGC~5448, however, no
ring-like feature is found.}
\label{fig:residual}
\end{figure}

Some of the observational effects of the asymmetrical dust
distribution could be interpreted as signatures of lopsidedness.
Lopsidedness in galaxies has been investigated by 
e.g., Baldwin, Lynden-Bell \& Sancisi (1980) and Swaters \etal (1999).
The amplitude of this effect may depend on galactocentric radius, 
and viewed from different viewing angles will produce different 
signatures in the observed velocity field. These studies have shown 
that the residuals are dominated by the zeroth harmonic term for 
viewing angle of $90\deg$, and by the second harmonic term for viewing 
angle of $0\deg$. In the special case of radius-dependent lopsidedness 
$c_0 \sim 3c_2$. NGC~5448 has a large inclination, thus the zeroth term 
dominates, and we find no correlation between the $c_0$ and the $c_2$ terms.
Our harmonic decomposition results display strongly varying zeroth and 
second terms. This was shown by SFdZ to be a poassible indication of an $m=1$ 
perturbation, i.e. lopsidedness (c.f., Fig 2. in SFdZ). 
In this case, subtracting the rotational component from the perturbed 
velocity field, one should find a ring-like feature. In Fig.~\ref{fig:residual}, 
we present this residual field for NGC~5448 and find no ring-like feature. 
Therefore dust is most likely the cause for the asymmetric features
that we observe in NGC~5448.

\subsection{The Large-scale Bar}
\label{sec:largebar} 
In Fig~\ref{fig:analysis}, we find that the harmonic parameters 
derived from the data are consistent with the bar model outlined 
in section \ref{sec:N5448bar}. Our simple model is limited to an 
$m=2$ perturbation. Complicated flow patterns in bars studied 
by, e.g., Lindblad (1999), imply a significant contribution from $m=4$ and 
higher-order modes. The third harmonic terms are also influenced by $m=4$, 
which could explain the differences between our model bar and the 
data for these parameters, but a detailed model is beyond the scope 
of this paper.

\begin{figure}
\center\includegraphics[scale=0.45,trim=0.cm 0.cm 0cm 1cm, angle=90]{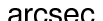} 
\caption{Observed (left) and deprojected (right) DSS image
of NGC~5448. The inclination of $64\deg$ and PA of $115\deg$ from
the RC3 catalogue have been used to deproject the image according
to the two-dimensional deprojection of a circular thin disk.}
\label{fig:DSSimage}
\end{figure}

We now compare the modelled bar with the deprojected Digitised Sky Survey 
(DSS) image. The inclination from the RC3 and the PA from our tilted-ring 
model were used to deproject the DSS image in a simple way by assuming that 
the galaxy disk is thin. The outer spiral arms are clearly emphasized in 
the deprojected DSS image (Fig.~\ref{fig:DSSimage}). 

It is somewhat difficult to identify the size of the large-scale
bar from the image. From our bar model we find a CR of about $37\arcsec$. 
In practice, this is done by assuming that at CR the circular frequency 
of the axisymmetric potential is identical to bar angular frequency. 
According to this definition and since we use a flat rotation curve, 
the ILR is located at $1 - {1/\sqrt{2}}$ of the CR, 
i.e., at about $11\arcsec$. If we assume that 
the bar in NGC~5448 ends close to the starting point of the 
spiral arms (Sanders \& Huntley 1976), we can also then 
associate its CR with the inner radius of the arms. 
Fig.~\ref{fig:DSSimage} shows that our analytic bar 
model is of reasonable size. The assumed CR radius 
is confirmed by the ellipticity profile presented 
in Fig.~\ref{fig:nicmos}, since the high ellipticity 
plateau starts at around this radius. At around the 
ILR (i.e., $11\arcsec$), there is a good agreement 
between the \sone\ profiles from the data and the 
model, with in both cases a change in sign at the ILR.

\section{Discussion and Conclusions}
\label{sec:conclusions}
We have used linear theory to construct a bar model ($m=2$ mode) for NGC~5448. 
The best-fit model considers the effects of a single weak bar with a potential 
ellipticity $\varepsilon = 0.15$ and a damping amplitude of $\lambda = 0.12$.
We compare the harmonic terms of the model with those derived from the data 
and find an overall consistency (see Fig.~\ref{fig:analysis}). 
Using this model, we have been able to associate the radial motion 
of gas with that of the large-scale bar (see Fig.~\ref{fig:s3s1}).

NGC~5448 exhibits clear signatures of the presence of other components 
than a single bar, which affect the observed velocity field. 
Inspecting the photometry and the central parts of the gaseous and stellar 
velocity field, we have detected a central rotating disk-like stellar 
component embedded in the larger disk (see Fig.~\ref{fig:diskmodel}). 
The WFPC2 image shows one very sharp dust lane 1\arcsec\ from 
the centre of this galaxy. This image also shows three more fuzzy and almost 
parallel dust lanes further out in the south-east direction. 
The dusty centre is also apparent in the $V-H$ image in Fig.~\ref{fig:nicmos}.
Dust is inhomogeneously distributed and continues down to the very bright 
nucleus. Light from the central light source is absorbed by the dust asymmetrically 
and the centre appears to be located at the north-west of its actual position. 

Investigating the projected and deprojected ellipticity profiles derived 
from the $H$-band data from Laine \etal (2002) confirms the presence of a 
central disk in the inner 7\arcsec. Although the projected ellipticity 
profile decreases at $\sim 4\arcsec$ and outwards, we cannot 
pin down the exact size of the central disk. The observed ellipticity 
decrease could be partly caused by the strongly asymmetric dust distribution. 
The stellar kinematic maps show that the central disk rotates faster 
than the main disk, and our observed gas distribution and kinematics 
indicate that this central disk also hosts gas which rotates 
faster than its stellar counterpart. This is not unexpected, 
since due to its disspative nature, the gas is not slowed down 
by asymmetric drift. It is important to note that, in NGC~5448, the
gas velocities are well ordered but larger in magnitude than the stellar
velocities. However, Fig.~\ref{fig:sauzoom} indicates that this velocity
difference dies out at larger radii. 
The difference between the stellar and gas velocities can be easily 
explained as due to the fact that the stars are in a thicker structure 
(bulge) in the central regions, while at larger radii both the gas and 
the stars are in a flatter disk, and both move closer to the circular velocity. 

It is known that bars are efficient in transferring mass towards the inner 
regions of their host galaxies. The centrally concentrated matter may be 
able to form a central disk (Yuan \& Yen 2004), or an inner bar
(e.g., Maciejewski \& Sparke 2000, Englmaier \& Shlosman 2004). 
Our analysis has shown that the non-circular gas kinematics 
in NGC~5448 could be driven by the large-scale bar. The central disk could
then have been formed as a result of the gas accumulation at the centre.

To conclude, we have been able to analytically model the 
bar signatures in the AGN host NGC~5448. We have unveiled a central
disk, and have distinguished the effects of a lopsided
perturbation from strong dust features. NGC~5448 hosts
considerable amounts of dust which is asymmetrically distributed
all the way to the centre, resembling the ``Evil Eye'' galaxy
(Braun, Walterbos \& Kennicutt 1992). This
study shows the power of the harmonic decomposition formalism to
quantify non-circular motions in observed velocity fields, and we
plan to apply this approach to the full set of
bulges observed with \Sauron.

\section*{Acknowledgements}
It is a pleasure to acknowledge the entire \Sauron\ team for their 
efforts in carrying out the observations and data preparation, for 
lively interactions and many fruitful discussions. 
The \Sauron\ project is made possible through grants
614.13.003 and 781.74.203 from ASTRON/NWO and financial contributions
from the Institut National des Sciences de l'Univers, the Universit\'e
Claude Bernard Lyon~I, the universities of Durham and Leiden, the
British Council, PPARC grant `Extragalactic Astronomy \& Cosmology at
Durham 1998--2002', and the Netherlands Research School for Astronomy
NOVA. Jes\'us Falc\'on-Barroso acknowledges support from the Euro3D 
Research Training Network. Kambiz Fathi acknowledges support for Proposal 
number HST-GO 09782.01 provided by NASA through a grant from the STScI, 
which is operated by the Association of Universities for Research in 
Astronomy, Incorporated, under NASA contract NAS5-26555. 
Kambiz Fathi is also grateful to Isaac Shlosman 
for stimulating and insightful discussions.
Finally, we thank the referee for constructive 
comments which improved the manuscript.


\label{lastpage}

\end{document}